Effects of ligand binding upon flexibility of proteins

Burak Erman

Chemical and Biological Engineering, Koc University, Istanbul, Turkey

Corresponding Author: Burak Erman

Email: berman@ku.edu.tr

Keywords:

Allostery, Flexibility, Dimerization, Stiffness Map, Gaussian Network Model, AKT1

ABSTRACT

Binding of a ligand on a protein changes the flexibility of certain parts of the protein, which directly affects its function. These changes are not the same at each point, some parts become more flexible and some others become stiffer. Here, an equation is derived that gives the stiffness map for proteins. The model is based on correlations of fluctuations of pairs of points that need to be evaluated by molecular dynamics simulations. The model is also cast in terms of the Gaussian Network Model and changes of stiffness upon dimerization of AKT1 are evaluated as an example.

INTRODUCTION

A major problem in protein physics is to understand how a local perturbation in a protein propagates through the structure over long distances. In an isotropic homogeneous linearly elastic solid, this problem has a unique and exact solution. Proteins are however neither homogeneous nor isotropic and the answer to this problem should be searched through a different route. In order to simplify the problem we compare the behavior of a linear elastic rod and a linear rod-like protein. The change in the length of the elastic rod is described by Hooke's law, $F = k\frac{\Delta L}{L}$ where F is the force, k is the spring constant, $\Delta L$ is the change in the length of the rod and L is the length. For a protein, on the other hand, the relationship between the force and the displacement comes through the Onsager relation [1], $\langle \Delta R^i \Delta R^j \rangle = kT\frac{\partial R^i}{\partial F^j}$ where, $\Delta R^i$ is the instantaneous fluctuation of one end of the rod, $\Delta R^j$ is the fluctuation of the other end, k is the Boltzmann constant, T is the absolute temperature, $R^i$ is the time averaged position of the end i, and $F^j$ is the force at the other end. Angular brackets denote the time average which denotes here the correlation between the fluctuation of one end relative to that of the other. Generalizing this problem to the protein, we see that perturbations in proteins result from the correlation of fluctuations of points in the protein, the point being the atom at the all-atom representation or the residue in the coarse grained picture. The aim of the present paper is to find an answer to the question: How is a point i displaced when a force acts at another point j. Answering this question sheds light on the structure-function relation for the protein which rests on evaluating the correlation of fluctuations of atoms.

Using the general approach set in the preceding paragraph, we focus on the effects produced at a point j resulting from the binding of a ligand at a point i. The 'ligand' may be a small molecule or atoms of another protein that docks into the protein of interest. In the derivation below, we first give the solution for the all-atom case and then simplify the problem to the coarse grained case described by the Gaussian Network Model, GNM.

MATERIALS AND METHODS

A quantitative measure of stiffness

The Onsager expression mentioned in the introduction can be written in terms of the Cartesian components of atoms as:

$$\langle \Delta R^i_m \Delta R^j_n \rangle = kT\frac{\partial R^i_m}{\partial F^j_n} \qquad (1)$$

where, the superscripts identify the residues and subscripts identify the Cartesian components. The Cartesian coordinates, m and n in Eq. 1 will be identified as 1, 2 and 3 denoting the x, y and z coordinates, respectively. $\langle \Delta R^i_m \Delta R^j_n \rangle$ is the correlation of m-direction fluctuations of atom i and n-direction fluctuations of j, $R^i_m$ is the m-component of the mean position of atom i and $F^j_n$ is the component of the force along n-direction acting on atom j. Now, assume we apply a force $F^j_n$ on atom j along the coordinate direction n. There may be forces on several atoms, each identified by the superscript j. We multiply both sides of Eq. 1 by $F^j_n$ and then first sum over n and then over j:

$$\sum_j \sum_n \langle \Delta R^i_m \Delta R^j_n \rangle dF^j_n = kT \sum_j \sum_n \frac{\partial R^i_m}{\partial F^j_n} dF^j_n$$

The summation is carried over all atoms on which the force is applied. But the right hand side of this equation is $(kT) dR^i_m$. Thus:

$$dR^i_m = \frac{1}{kT} \sum_j \sum_n \langle \Delta R^i_m \Delta R^j_n \rangle dF^j_n \qquad (2)$$

$dR^i_m$ is the displacement of residue i along the m-direction. Equation 2 is a simple demonstration of the linear response theory.

Let us now take two alpha carbons, one of residue i, the other of residue j. Let us apply equal and opposite forces of unit magnitude to atoms i and j, a force $F^j_n$ on atom j and a force $F^i_n = -F^j_n$ on atom i. The m-component of the displacements of atoms j and i under the applied forces are

$$dR^j_m = \sum_n \left[ \langle \Delta R^j_m \Delta R^j_n \rangle - \langle \Delta R^j_m \Delta R^i_n \rangle \right] F^j_n$$

$$dR^i_m = -\sum_n \left[ \langle \Delta R^i_m \Delta R^i_n \rangle - \langle \Delta R^i_m \Delta R^j_n \rangle \right] F^j_n \qquad (3)$$

The choice of alpha carbons is for simplicity, the derivation is valid for any atom. The m-component of the relative change of the positions of atoms i and j along the direction from atom i to atom j due to a unit tensile force that lies along the line connecting i and j is

$$\delta R^{ij}_m = d\left(R^j_m - R^i_m\right) u^{ij}_m = d\left(R^j_m - R^i_m\right) u^{ij}_m =$$
$$\sum_n \left[ \langle \Delta R^j_m \Delta R^j_n \rangle - \langle \Delta R^j_m \Delta R^i_n \rangle - \langle \Delta R^i_m \Delta R^j_n \rangle + \langle \Delta R^i_m \Delta R^i_n \rangle \right] u^{ij}_m u^{ij}_n$$

(4)

where, $\delta R^{ij}_m$ is the mth component of the change of the vector from i to j, and $u^{ij}_m$ is the m'th component of the unit vector from atom i to atom j, given by

$u^{ij}_m = \frac{\left(R^j_m - R^j_m\right)}{\left|R^j - R^i\right|}$. The second unit vector on the right hand side of Eq. 4 comes from using a unit force that lies along the ij direction. The two vertical lines in the denominator here denote the magnitude of the quantity inside them.

The change in length per length of the distance from i to j is $\dfrac{\left[\sum_m \left(\delta R^{ij}_m\right)^2\right]^{0.5}}{\left|R^j - R^i\right|}$

Substituting from Eq 4 leads to the change in length per unit length, which is inverse of the spring constant between i and j

$$k_{ij}^{-1} = \left\{\sum_m\left[\sum_n\left(\langle\Delta R^j_m\Delta R^j_n\rangle - \langle\Delta R^j_m\Delta R^i_n\rangle - \langle\Delta R^i_m\Delta R^j_n\rangle + \langle\Delta R^i_m\Delta R^i_n\rangle\right)u^{ij}_m u^{ij}_n\right]^2\right\}^{0.5} / \left|R^j - R^i\right| \quad (5)$$

The spring constant $k_{ij}$, a fictitious entity, is a measure of the rigidity of the distance between residues i and j. It is now possible to construct a stiffness map, similar to the contact map, for the protein. However, the differences of stiffness maps between the liganded and free proteins is more instructive. Finding the value of $k_{ij}$ for a free protein and for a protein on which a ligand is bound gives us the stiffening effects of ligand binding on proteins. Thus, we need to evaluate the correlation of fluctuations twice, once for the free and once for the ligand bound protein. A convenient way of obtaining the correlations at the all-atom scale is by molecular dynamics simulations. Thus, one has to run two sets of simulations, calculate $k_{ij}$ from Eq. 5 and compare.

The derivations presented here may be simplified by assuming that interactions are at a coarse grained level and are isotropic, as a result of which only residues in contact contribute and the problem reduces to that of the Gaussian Network Model (GNM).

The Gaussian Network Model

In the GNM [2], two residues interact only if they are within a contact distance of each other, hence only inner products are considered. The correlation $\langle\Delta R^i_m \Delta R^j_n\rangle$ in Eq. 1 becomes the inner (dot) product of two vectors, $\Delta \boldsymbol{R}^i$ and $\Delta \boldsymbol{R}^j$, as a result of which Eq. 1 reduces to

$$\sum_{m,n}\langle\Delta R^i_m \Delta R^j_n\rangle \delta_{mn} = kT \sum_{m,n} \frac{\partial R^i_m}{\partial F^j_n}\delta_{mn} \quad (6)$$

where $\delta_{mn}$ is the Kronecker delta symbol. Equation 6 takes the more concise form

$$\langle\Delta R^i \Delta R^j\rangle = kT\frac{\partial R^i}{\partial F^j} \quad (7)$$

where the right hand side is the sum of the gradients in three directions

$$\frac{\partial R^i}{\partial F^j} = \frac{\partial R^i_x}{\partial F^j_x} + \frac{\partial R^i_y}{\partial F^j_y} + \frac{\partial R^i_z}{\partial F^j_z} \quad (8)$$

Writing the GNM expression, $\Delta \boldsymbol{F}^i = \Gamma^{ij}\Delta \boldsymbol{R}^j$ in inverse form, $\Delta \boldsymbol{R}^i = \left(\Gamma^{-1}\right)^{ij}\Delta \boldsymbol{F}^i$ and differentiating, we obtain

$$\frac{\partial R^i}{\partial F^j} = \left(\Gamma^{-1}\right)^{ij} \tag{9}$$

Comparing Eqs. 7 and 9 shows that the inverse of the gamma matrix of GNM is the correlation matrix,

$$\left\langle \Delta R^i \Delta R^j \right\rangle = kT \left(\Gamma^{-1}\right)^{ij} \tag{10}$$

The matrix $\Gamma$ is obtained from the residue based coarse grained picture of the protein as follows: If there is a contact between residues i and j, contact being assumed if the two residues are within a priori given cutoff distance which may plausibly taken as the first coordination shell of a residue, then the ij'th element of $\Gamma$ is the spring constant. The diagonal elements of $\Gamma$ equals to the negative sum of the corresponding rows [2].

Equation 5 takes the following form for the GNM

$$k_{ij}^{-1} = \left(\left(\Gamma^{-1}\right)^{jj} - 2\left(\Gamma^{-1}\right)^{ij} + \left(\Gamma^{-1}\right)^{ii}\right) / \left|R^j - R^i\right| \tag{11}$$

RESULTS AND DISCUSSION

Change of flexibility of ACK1 upon dimerization by GNM

The activated Cdc42-associated kinase-1, abbreviated as ACK1, is a tyrosine kinase that is involved in the regulation of cell adhesion and growth, receptor degradation, and axonal guidance. It is also involved in several different signaling pathways [3]. ACK1 is activated allosterically, where ligand binding at different locations induce different conformational changes in AKT1, resulting in activation or inactivation, reviewed in Reference [4]. Here, we compare flexibility changes upon two different binding conformations. In the first case, we study the PDB structure 4HZR from Reference [3] where AKT1 is activated upon dimerization, and in the second we study 1U46 from Reference [5], which is the same protein as 4HZR but in a different conformation. The two cases are shown in Figure 1 for 4HZR (left) and 1U46 (right).

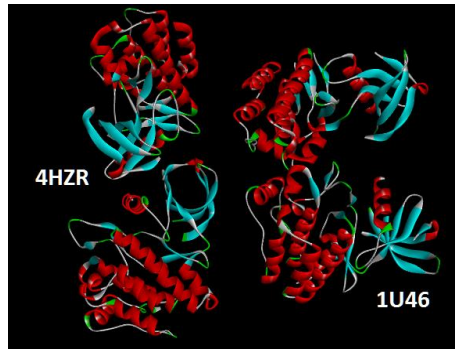

Figure 1. Dimerization conformations of 4HZR and 1U46.

We used the GNM in order to see changes in the stiffness of different parts of AKT1 upon binding. The cutoff distance for contacts is taken as 7.0 Å and Eq 11 is used for the stiffness of the distance between residues i and j. The difference between stiffness maps of 4HZR, which is the active dimer, is shown in Figure 2. It is obtained as follows: First the $\Gamma$ matrix of AKT1 as a monomer is obtained. Then, the residues of the second monomer that are within 7Å distance of those of the first monomer are added to the $\Gamma$ matrix. Stiffness maps are obtained from Eq 11 with each $\Gamma$ matrix and the differences of $k_{ij}(\text{dimer}) - k_{ij}(\text{monomer})$ are shown in Figure 2. The left panel shows the positive values of this difference indicating the increase of stifness and the right panel shows the decrease. Figure 3 for 1U46, the inactive case, is obtained similarly.

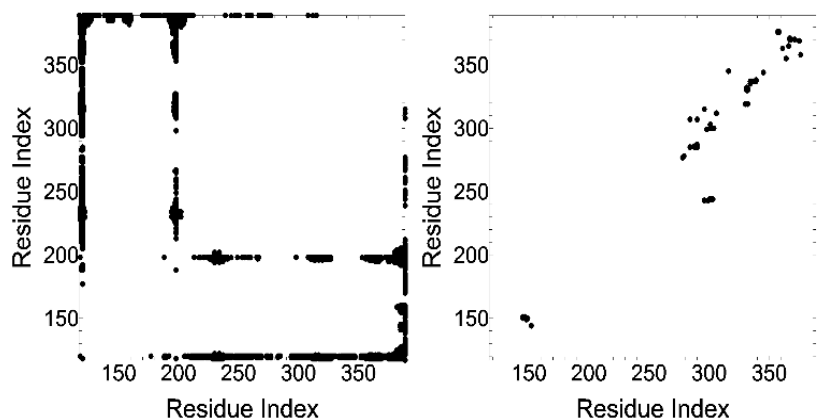

Figure 2. Increase of stiffness upon dimerization (left), decrease (right) for 4HZR (the active dimer)

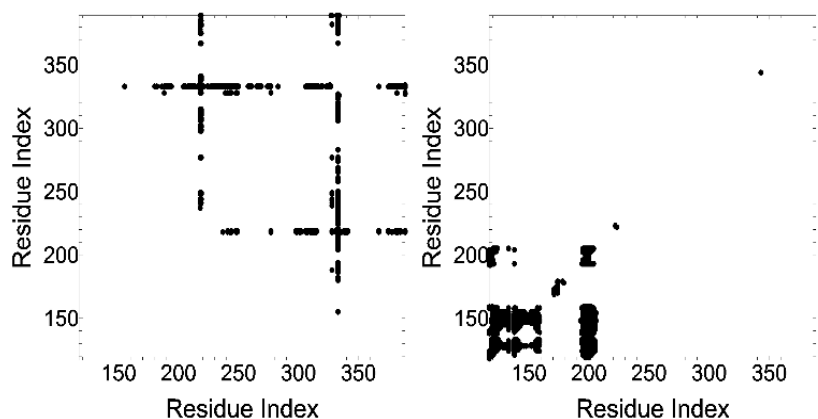

Figure 3. Increase of stiffness upon dimerization (left), decrease (right) for 1U46 (the inactive dimer).

Figures 2 and 3 show that the changes in flexibility upon binding depend on the mode of binding and that these changes are not uniform, some parts becoming stiffer while to others becoming more flexible. The amplitudes of changes are not shown on the figures, but in all cases, stiffening is more dominant. The GNM is only a coarse model and one should be warned that Figures 2 and 3 may only be qualitative. However, these results conclusively show, in the proof of concept manner, the dependence of protein flexibility upon ligand binding. A quantitatively reliable evaluation will be possible only by the use of Eq. 5 and extensive and careful molecular dynamics simulations.